\documentclass[twocolumn,showpacs,aps,prl,
groupedaddress,amssymb,amsmath,nobalancelastpage]{revtex4}

\usepackage{graphicx}
\usepackage{longtable}

\begin{document}

\title{Reversible plasticity in amorphous materials }

\author{Micah Lundberg$^1$, Kapilanjan Krishan$^1$, Ning Xu$^{2,3}$, Corey S. O'Hern$^{4,5}$ and Michael Dennin$^1$}
\affiliation{$^1$Department of Physics and Astronomy, University
of California at Irvine, Irvine, CA 92697-4575\\
$^{2}$Department of Physics and Astronomy, University of
Pennsylvania, Philadelphia, PA 19104-6396\\
$^{3}$James Franck Institute, The University of Chicago, Chicago,
IL 60637\\
$^4$Department of Mechanical Engineering, Yale University, New Haven,
CT 06520-8286\\
$^5$Department of Physics, Yale University, New Haven, CT 06520-8120}
\date{\today}

\begin{abstract}
A fundamental assumption in our understanding of material rheology is
that when microscopic deformations are reversible, the material
responds elastically to external loads. Plasticity, i.e.  dissipative
and irreversible macroscopic changes in a material, is assumed to be
the consequence of irreversible microscopic events.  Here we show
direct evidence for reversible plastic events at the microscopic scale
in both experiments and simulations of two-dimensional foam. In the
simulations, we demonstrate a link between reversible plastic
rearrangement events and pathways in the potential energy landscape of
the system. These findings represent a fundamental change in our
understanding of materials---microscopic reversibility does not
necessarily imply elasticity.
\end{abstract}

\maketitle

One of the fundamental questions in materials science concerns the
microscopic origin of plastic behavior.  Why do materials display
plastic rather than elastic and reversible response and can we predict
for what loads this will occur?  An improved understanding of plastic
deformation is especially important in a wide range of amorphous
materials, such as metallic\cite{A79,FLP04} and polymeric
glasses\cite{E00}, viscoplastic solids\cite{B42}, foams\cite{WH99},
granular materials\cite{LN93}, colloids\cite{WCLSW00,WW02},
emulsions\cite{MBW96}, and even intracellular networks\cite{HLM03}. In
crystalline materials, plastic behavior is understood in terms of
defect nucleation and dynamics\cite{RT74,R92}. However, for amorphous
materials, a description in terms of topological defects is not
possible due to inherent structural disorder. Therefore, identifying
and characterizing local plastic events in amorphous materials is
essential for a complete understanding of their structural and
mechanical properties. The conventional wisdom is that plastic
rearrangement events cause irreversible structural changes in these
materials on the microscale.

The macroscopic response of amorphous solids and complex fluids,
such as foams, colloids, and granular matter, to applied stress
and strain is very similar: elastic at small strains and plastic
at larger strains. In the elastic regime, stress is proportional
to applied strain, and deformations are reversible. Above the
yield stress or strain, plastic flow or anelastic deformation
occurs. Given these similarities on the macroscopic scale, many
models of plasticity have emphasized the importance of
microscopic ``plastic zones" within amorphous
materials\cite{FLP04,BVR02,FL98,HSF05,L06,O03,PABL02} in which
neighbor switching and other rearrangements events of ``particles"
occur. (The particles represent molecules in the case of solids,
or bubbles or grains in the case of complex fluids.) An important
open question concerning plasticity is whether or not plastic
zones are intrinsically irreversible or, instead, is their
surrounding environment ultimately responsible for determining
whether or not rearrangement events are reversible?

We perform both experiments and simulations of two-dimensional
amorphous foams undergoing oscillatory shear strain to investigate
this fundamental question. In both cases, we find a significant
fraction of dissipative, plastic rearrangement events that are {\it
reversible}, even for strains significantly above the yield strain. In
the simulations, measurements of the local potential energy allows us
to assess the impact of the bubble's neighborhood on the reversibility
of the plastic events. This links reversible plastic rearrangement
events to pathways in the potential energy landscape of the system
during deformation. We argue that even during plastic flow certain
microscopic rearrangement events are intrinsically reversible and
changes in the environment surrounding plastic zones determine whether
the zones are reversible or not.

We chose bubble rafts\cite{B42,AK79,KE99,D04} that consist of gas
bubbles floating on a water surface for our experimental system. For
the simulations, we employed the well-characterized bubble model for
two-dimensional (2D) foams developed by Durian\cite{D95}.  The bubble
model assumes massless circular bubbles that interact through a
repulsive linear spring force and viscous dissipation.  Experimental
evidence supports the applicability of the bubble model to explain the
flow behavior of bubble rafts, as well as three-dimensional
foam\cite{LTD02,D04,GD95}. Even though other rearrangement events
occur in bubble rafts and the bubble model, plastic rearrangement
events known as T1 events play a central role in the mechanical
response of
foam\cite{D04,WKD07,DK97,DDG06,KE99,RK00,VC05,WBHA92,VHC06}.

T1 events correspond to a neighbor switching event in which two
neighboring bubbles lose contact, and two next-nearest neighbors
become neighbors\cite{WH99}. This corresponds to a transition between
two distinct states of the system. For example, referring to Fig. 1,
State A is when bubbles 1 and 2 are neighbors, and State B is when
bubbles 3 and 4 are neighbors. For both the experiment and the
simulations, during one cycle the applied shear strain varies from $0$
to $A/L_y$ (at phase $\psi = \pi$) and back to a strain of $0$ (at
$\psi = 2\pi$), where $A$ is the amplitude of the shear displacement
and $L_y$ is the system size in the shear-gradient direction. If four
bubbles experience a T1 event that switched the bubbles from state A
to B during the first half-cycle of the drive, a reversible T1 event
occurs if the same foursome of bubbles returns to state A in the
second half-cycle of the drive. Otherwise, the T1 event is
irreversible.

For the experiments, the system contained approximately $800$
bubbles in a planar shear cell with $L_y = 9\ {\rm cm}$.  Half of
the bubbles were 2~mm in diameter and the other half were 3.5~mm.
We report on results using driving amplitudes $A$ of 10 and 12
times the diameter of the small bubbles and driving frequency
$0.2\ {\rm s^{-1}}$. The resulting rms strain and strain rate were
approximately $0.2$ and $0.04\ {\rm s^{-1}}$, respectively. This
should be compared with the yield strain of $0.01$ for bubble
rafts and the transition to quasi-static behavior on the order of
$0.07\ {\rm s^{-1}}$\cite{pratt}. At the yield strain, T1 events
give rise to permanent plastic deformation\cite{H97}.  Details of
the experimental setup for bubble rafts can be found in
Ref.~\cite{lkxod07b}.

In the bubble model simulations, bidisperse systems composed of $N/2$
large and $N/2$ small circular bubbles with diameter ratio $r=1.75$
were used to match experiments.  We studied square simulation cells
with system sizes in the range $N = 16$ to $1024$ and packing fraction
$\phi = 0.95$, so that the bubbles were always compressed during
shear.  The bubble model treats foams as massless deformable disks
with an equation of motion that balances a linear repulsive spring
force to model elastic repulsion with viscous dissipation proportional
to local velocity differences\cite{D95}.  The oscillatory shear strain
is applied quasistatically to the system by shifting the x-positions
of the bubbles, implementing shear-periodic Lees-Edwards boundary
conditions\cite{allen} and minimizing the total potential energy.  To
study the role of the energy landscape, two definitions of the local
potential energy based on the overlaps between bubbles were used, $E$
and $E'$. $E$ is computed only considering overlaps among the four
bubbles defining the T1 event, while $E'$ also includes overlaps
with the first nearest neighbors of the T1 bubbles. Finally, we
measured $\Delta E'$, defined by subtracting the potential energy $E'$
of the four bubbles participating in the reversible T1 event from the
original oscillatory shear strain simulations with $E'$ from
simulations in which the four T1 bubbles are forced to exactly retrace
their positions as they transition from state B back to state A, but
all other particles are allowed to move without constraints.

\begin{figure}
\includegraphics[width=8.6cm]{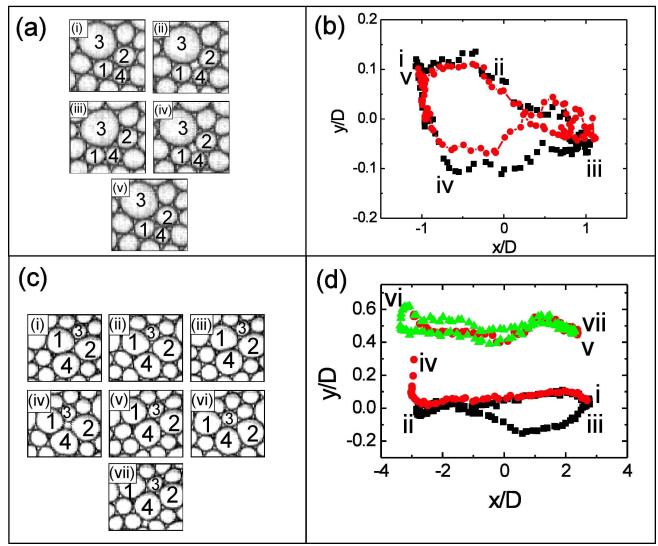}
\caption{(color online) Experimental results for a reversible [(a)
and (b)] and an irreversible T1 event [(c) and (d)].  The images
in (a) and (c) are $10.5~{\rm mm} \times 10.5~{\rm mm}$ with
bubbles involved in the T1 events labeled by numbers.  The roman
label for each image in (a) and (c) corresponds to the same label
on the trajectory in (b) and (d). (a) The images highlight the
following stages of the reversible T1 event: initial state (i),
middle of the T1 event (ii), second state (iii), middle of the T1
event under reversal of shear (iv), and the return to the initial
state (v)\cite{EPAPS}. (b) Plot of the trajectory of bubble 1 in
the images in (a). Two consecutive cycles are shown (the first in
black squares and the second in red circles). (c) The images
highlight the following stages of the irreversible T1 event:
initial state (i - iii), middle of the T1 event (iv), and the
second state (v - vii). (d) Plot of the trajectory of bubble 4 in
the images in (c). Three consecutive cycles are shown (the first
in black, the second in red, the third in green).} \vspace{-0.2in}
\end{figure}

\begin{figure}
\includegraphics[width=8.6cm]{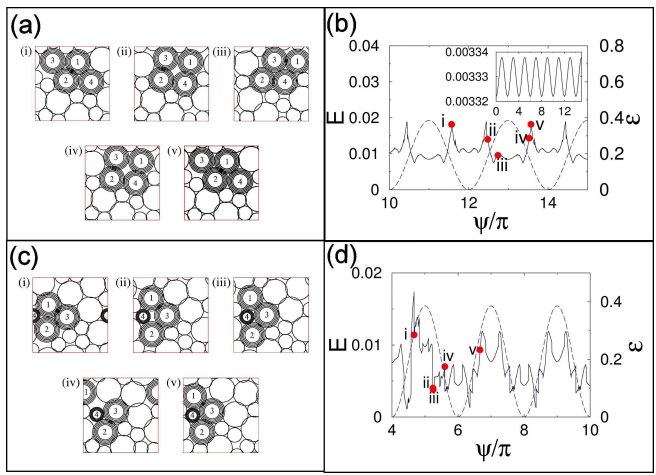}
\caption{ Results taken from a 16-particle simulation of the
bubble model in two dimensions undergoing oscillatory shear strain
with an amplitude of $2$ small bubble diameters for a reversible
[(a) and (b)] and an irreversible event [(c) and (d)]. Bubbles
involved in the T1 events are labeled by numbers. Roman labels in
the images correspond to the same labels in the plots. (a) Images
(i)-(iii) show the occurrence of a reversible T1 event and (iv)
and (v) show the reversal of the T1 event. (See the supplementary
information for a movie of this event.)  (b) The local potential
energy $E$ (solid black line) is plotted versus the driving phase
(left axis). For comparison, the periodic strain is plotted with a
long dashed blue line (right axis). The elapsed phase for the T1
event is significantly different than that for the reversed T1
event, and the shape of the local potential energy is not the same
for the T1 event and its reverse. The inset shows an elastic
response in $E$ vs $\psi/\pi$ for small amplitude oscillations
$A=10^{-2}$ times the small bubble diameter, where the response
matches the driving. (c) Images (i)-(iii) show the occurrence of
an irreversible T1 event, as shown by the absence of the reverse
T1 event in images (iv) and (v)\cite{EPAPS}. (d) The local
potential energy $E$ (solid black line) is plotted versus the
driving phase (left axis). The periodic strain is plotted with a
long dashed blue line (right axis). Note that $E$ for locations
(i) and (v) separated by a phase interval of $2\pi$ are not the
same. \vspace{-0.3in} }

\end{figure}

To fully understand the behavior of the system, it is best to directly
compare the reversible and irreversible rearrangement events from both
experiments and simulation. Panels (a) and (b) of Fig.~1 (experiment)
and Fig.~2 (simulation) highlight typical reversible T1 events. Panels
(c) and (d) of Fig.~1 (experiment) and Fig.~2 (simulation) highlight
irreversible events. The plots focus on the four bubbles (labeled 1 -
4) that experience a T1 event.  Snapshots are used to illustrate
bubble motions during a typical T1 event (panels (a) and (c)). For the
experiments, panel (b) and (d) highlight the trajectory of a single
bubble in real space. For the simulations, panels (b) and (d) display
the potential energy $E$ as a function of the phase of the
driving. The simulation results are for $N=16$, but similar results
were obtained with much larger systems with $N=1024$.

The defining feature of reversible T1 events is that the initial and
final states of the bubbles in the T1 event are equivalent,
despite the occurrence of dissipation. The
dissipation is evident in the anharmonic behavior of the local
potential energy signal. This is in contrast to the perfectly
elastic behavior for small shear strains in the absence of T1
events shown in the inset to Fig.~2b. The existence of the
dissipation leads to a number of asymmetries in the dynamics of
the system, despite the overall periodic nature of the response.
The spatial trajectory is a closed loop with a finite area
(see Fig.~1b). This causes the symmetric rearrangements of the four
bubbles during the two T1 events (images (ii) and (iv) in Fig.~1a)
to occur at different locations during the corresponding
half-cycle. Likewise, Fig.~2b illustrates that $E$ is very
different for the two half cycles corresponding to labels
(i)-(iii) and (iv)-(v). Finally, the durations of the T1 event and
its reverse (state A to B vs. state B to A) are not the same.

During irreversible T1 events, the defining feature is that a
foursome of bubbles undergoes a T1 event in the first half cycle
of the driving but the reverse T1 event does not occur during the
second half cycle. The experimental example in Fig.~1c is a case
where a single T1 event from state A to B occurred during seven
cycles of the driving (three of which are highlighted in Fig.~1c).
The impact of the T1 event on the trajectories in Fig.~1d is
dramatic. The trajectory of bubble four is shown for three cycles:
just before the T1 event (black), during the T1 event (red), and
just after the T1 event (green). In the absence of a T1 event, the
local trajectory essentially repeats itself during each half-cycle
as the bubbles move along similar paths. The occurrence of the T1
event represents a dramatic break in this motion.

The behavior of irreversible T1 events in simulations is similar
to that found in experiments. Figure 2c illustrates a T1 event
that occurs during frames (i)-(iii) in the first half cycle, but
the T1 event is not reversed during the second half cycle in
frames (iii)-(v). The plot of $E$ in Fig.~2(d) illustrates that
configurations (i) and (v), which are separated by $2\pi$ in
phase, do not have the same local potential energy. Note that
beyond $\psi/\pi \sim 5.5$, the local potential energy signal is
periodic, which indicates that other T1 events or possibly more
complex rearrangement events that occur in the system are
reversible.

\begin{figure}
\scalebox{0.35}{\includegraphics{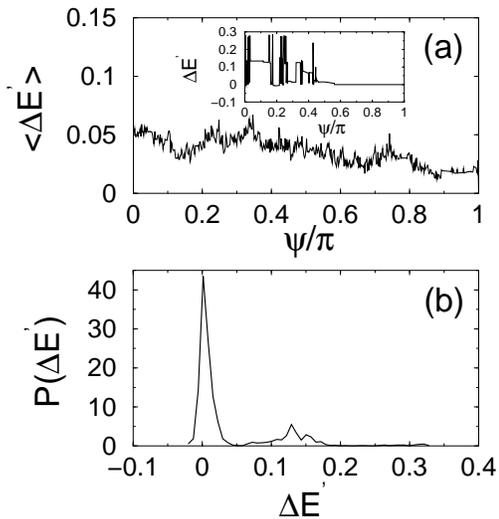}}
\vspace{-0.2in}
\caption{(a) The local potential energy difference $\langle \Delta E'
\rangle$ averaged over $100$ reversible T1 events plotted vs. the
driving phase under the same conditions in Fig.~2.  $\langle \Delta E'
\rangle > 0$ confirms that {\em exact} trajectory reversal is not
energetically favorable. The inset shows $\Delta E'$ for a single
reversible T1 event. (b) The probability of finding a particular
$\Delta E'$. There is a large peak at $\Delta E' = 0$, two slight
peaks near $0.15$ and $0.30$, and no significant weight in the
distribution for $\Delta E' < 0$.
\vspace{-0.25in}
}
\end{figure}

What is the connection between reversible T1 events and the
path that the system follows through the potential energy
landscape? Answering this question provides initial insights into
why some T1 events are reversible and others are irreversible and
how the system returns to the same potential energy minimum even
though it follows a different path in the energy landscape during
the T1 event and its reverse. The calculation of $\Delta
E'$ obtained by comparing the local potential energy
$E'$ (including interactions of T1 bubbles with first
nearest neighbors) of the four bubbles in the original oscillatory
shear strain simulations with $E'$ from the constrained
simulations directly addresses this question.

The results from these studies are shown in Fig.~3. First, in
Fig.~3(a) we find that the local potential energy difference
averaged over many T1 events from independent runs $\langle \Delta
E' \rangle > 0$. $\Delta E'$ for a single T1 event shown in the
inset to Fig.~5(a) has large positive spikes but also significant
phase intervals where $\Delta E' = 0$. In Fig.~3(b), we show that
the distribution $P(\Delta E')$ of energy differences has a strong
peak at zero, but non-negligible peaks at $\Delta E^{\prime}
\approx 0.15$ and $0.3$ and no significant weight for $\Delta
E^{\prime} < 0$.  Each of these findings indicates that the path
during the second half cycle that does not exactly retrace the
path in configuration space of the first half cycle is
energetically favorable.  Furthermore, for the case of reversible
T1 events, it is likely that there are a number of local low
energy pathways that lead from state B back to state A. For the
case of irreversible T1 events, it is likely that there are many
local energy pathways away from state B, but the ones that are
energetically favorable do not lead back to state A. Since our
system is athermal, these differences in the energy pathways are
due to changes in the environment (surrounding bubbles) that occur
during the applied shear strain. Thus, we argue that the influence of
the environment gives rise to the irreversibility of T1 events in
foams. In equilibrium systems, thermal fluctuations will also play
a significant role in determining reversibility.

Our experiments and simulations of model foams undergoing
oscillatory shear strain identify reversible T1 events, which are
two state systems. This observation is the first direct
experimental confirmation of a general two-state model of
plasticity: shear transformation zones (STZ). The concept of a STZ
as a reversible, two-state transition within a material was first
proposed by Falk and Langer\cite{FL98}.  The STZ picture is
successful in explaining a range of macroscopic behavior of
materials based on dynamics of the microstructure. STZ's represent
a natural extension of ideas based on activated transitions and
free volume\cite{A79,KK94,S77} and it has motivated a number of
other models of plasticity\cite{PABL02,B03,PALB04}. Therefore, our
results establish the applicability of two-state STZ models to
athermal particulate systems, and the need to include
intrinsically reversible plastic events in models of plasticity.
Our studies of the local potential energy landscape go beyond the
two-state model and establish the importance of the accessible
pathways in the energy landscape that ultimately determine the
reversibility of the plastic events. Thus, we have learned that
plasticity does not imply microscopic irreversibility and that
microscopic reversibility does not imply elasticity.

\begin{acknowledgments}

Financial support from the Department of Energy grant numbers
DE-FG02-03ED46071 (MD), DE-FG02-05ER46199 (NX), and DE-FG02-03ER46088
(NX), NSF grant numbers DMR-0448838 (CSO, GL) and CBET-0625149 (CSO),
and the Institute for Complex Adaptive Matter (KK) is gratefully
acknowledged.  We also thank M. Falk for insightful conversations.
\end{acknowledgments}


\begin{thebibliography}{40}
\expandafter\ifx\csname
natexlab\endcsname\relax\def\natexlab#1{#1}\fi
\expandafter\ifx\csname bibnamefont\endcsname\relax
  \def\bibnamefont#1{#1}\fi
\expandafter\ifx\csname bibfnamefont\endcsname\relax
  \def\bibfnamefont#1{#1}\fi
\expandafter\ifx\csname citenamefont\endcsname\relax
  \def\citenamefont#1{#1}\fi
\expandafter\ifx\csname url\endcsname\relax
  \def\url#1{\texttt{#1}}\fi
\expandafter\ifx\csname
urlprefix\endcsname\relax\def\urlprefix{URL }\fi
\providecommand{\bibinfo}[2]{#2}
\providecommand{\eprint}[2][]{\url{#2}}

\bibitem[{\citenamefont{Argon}(1979)}]{A79}
\bibinfo{author}{\bibfnamefont{A.~S.} \bibnamefont{Argon}},
  \bibinfo{journal}{Acta Metallurgica} \textbf{\bibinfo{volume}{27}},
  \bibinfo{pages}{47} (\bibinfo{year}{1979}).

\bibitem[{\citenamefont{Falk et~al.}(2004)\citenamefont{Falk, Langer, and
  Pechenik}}]{FLP04}
\bibinfo{author}{\bibfnamefont{M.~L.} \bibnamefont{Falk}},
  \bibinfo{author}{\bibfnamefont{J.~S.} \bibnamefont{Langer}},
  \bibnamefont{and} \bibinfo{author}{\bibfnamefont{L.}~\bibnamefont{Pechenik}},
  \bibinfo{journal}{Physical Review E} \textbf{\bibinfo{volume}{70}},
  \bibinfo{pages}{011507} (\bibinfo{year}{2004}).

\bibitem[{\citenamefont{Ediger}(2000)}]{E00}
\bibinfo{author}{\bibfnamefont{M.~D.} \bibnamefont{Ediger}},
  \bibinfo{journal}{Annual Review Of Physical Chemistry}
  \textbf{\bibinfo{volume}{51}}, \bibinfo{pages}{99} (\bibinfo{year}{2000}).

\bibitem[{\citenamefont{Bragg}(1942)}]{B42}
\bibinfo{author}{\bibfnamefont{L.}~\bibnamefont{Bragg}},
  \bibinfo{journal}{Journal of Scientific Instruments}
  \textbf{\bibinfo{volume}{19}}, \bibinfo{pages}{148} (\bibinfo{year}{1942}).

\bibitem[{\citenamefont{Weaire and Hutzler}(1999)}]{WH99}
\bibinfo{author}{\bibfnamefont{D.}~\bibnamefont{Weaire}} \bibnamefont{and}
  \bibinfo{author}{\bibfnamefont{S.}~\bibnamefont{Hutzler}},
  \emph{\bibinfo{title}{The Physics of Foams}} (\bibinfo{publisher}{Clarendon
  Press}, \bibinfo{address}{Oxford}, \bibinfo{year}{1999}).

\bibitem[{\citenamefont{Liu and Nagel}(1993)}]{LN93}
\bibinfo{author}{\bibfnamefont{C.~H.} \bibnamefont{Liu}} \bibnamefont{and}
  \bibinfo{author}{\bibfnamefont{S.~R.} \bibnamefont{Nagel}},
  \bibinfo{journal}{Phys. Rev. B} \textbf{\bibinfo{volume}{48}},
  \bibinfo{pages}{15646} (\bibinfo{year}{1993}).

\bibitem[{\citenamefont{Weeks et~al.}(2000)\citenamefont{Weeks, Crocker,
  Levitt, Schofield, and Weitz}}]{WCLSW00}
\bibinfo{author}{\bibfnamefont{E.~R.} \bibnamefont{Weeks}},
  \bibinfo{author}{\bibfnamefont{J.~C.} \bibnamefont{Crocker}},
  \bibinfo{author}{\bibfnamefont{A.~C.} \bibnamefont{Levitt}},
  \bibinfo{author}{\bibfnamefont{A.}~\bibnamefont{Schofield}},
  \bibnamefont{and} \bibinfo{author}{\bibfnamefont{D.~A.} \bibnamefont{Weitz}},
  \bibinfo{journal}{Science} \textbf{\bibinfo{volume}{287}},
  \bibinfo{pages}{627} (\bibinfo{year}{2000}).

\bibitem[{\citenamefont{Weeks and Weitz}(2002)}]{WW02}
\bibinfo{author}{\bibfnamefont{E.~R.} \bibnamefont{Weeks}} \bibnamefont{and}
  \bibinfo{author}{\bibfnamefont{D.~A.} \bibnamefont{Weitz}},
  \bibinfo{journal}{Physical Review Letters} \textbf{\bibinfo{volume}{89}},
  \bibinfo{pages}{095704} (\bibinfo{year}{2002}).

\bibitem[{\citenamefont{Mason et~al.}(1996)\citenamefont{Mason, Bibette, and
  Weitz}}]{MBW96}
\bibinfo{author}{\bibfnamefont{T.~G.} \bibnamefont{Mason}},
  \bibinfo{author}{\bibfnamefont{J.}~\bibnamefont{Bibette}}, \bibnamefont{and}
  \bibinfo{author}{\bibfnamefont{D.~A.} \bibnamefont{Weitz}},
  \bibinfo{journal}{Journal of Colloid and Interface Science}
  \textbf{\bibinfo{volume}{179}}, \bibinfo{pages}{439} (\bibinfo{year}{1996}).

\bibitem[{\citenamefont{Head et~al.}(2003)\citenamefont{Head, Levine, and
  MacKintosh}}]{HLM03}
\bibinfo{author}{\bibfnamefont{D.~A.} \bibnamefont{Head}},
  \bibinfo{author}{\bibfnamefont{A.~J.} \bibnamefont{Levine}},
  \bibnamefont{and} \bibinfo{author}{\bibfnamefont{F.~C.}
  \bibnamefont{MacKintosh}}, \bibinfo{journal}{Physical Review E}
  \textbf{\bibinfo{volume}{68}}, \bibinfo{pages}{061907}
  (\bibinfo{year}{2003}).

\bibitem[{\citenamefont{Rice and Thomson}(1974)}]{RT74}
\bibinfo{author}{\bibfnamefont{J.~R.} \bibnamefont{Rice}} \bibnamefont{and}
  \bibinfo{author}{\bibfnamefont{R.}~\bibnamefont{Thomson}},
  \bibinfo{journal}{Philosophical Magazine} \textbf{\bibinfo{volume}{29}},
  \bibinfo{pages}{73} (\bibinfo{year}{1974}).

\bibitem[{\citenamefont{Rice}(1992)}]{R92}
\bibinfo{author}{\bibfnamefont{J.~R.} \bibnamefont{Rice}},
  \bibinfo{journal}{Journal Of The Mechanics And Physics Of Solids}
  \textbf{\bibinfo{volume}{40}}, \bibinfo{pages}{239} (\bibinfo{year}{1992}).

\bibitem[{\citenamefont{Baret et~al.}(2002)\citenamefont{Baret, Vandembroucq,
  and Roux}}]{BVR02}
\bibinfo{author}{\bibfnamefont{J.~C.} \bibnamefont{Baret}},
  \bibinfo{author}{\bibfnamefont{D.}~\bibnamefont{Vandembroucq}},
  \bibnamefont{and} \bibinfo{author}{\bibfnamefont{S.}~\bibnamefont{Roux}},
  \bibinfo{journal}{Physical Review Letters} \textbf{\bibinfo{volume}{89}},
  \bibinfo{pages}{195506} (\bibinfo{year}{2002}).

\bibitem[{\citenamefont{Falk and Langer}(1998)}]{FL98}
\bibinfo{author}{\bibfnamefont{M.~L.} \bibnamefont{Falk}} \bibnamefont{and}
  \bibinfo{author}{\bibfnamefont{J.~S.} \bibnamefont{Langer}},
  \bibinfo{journal}{Physical Review E} \textbf{\bibinfo{volume}{57}},
  \bibinfo{pages}{7192} (\bibinfo{year}{1998}).

\bibitem[{\citenamefont{Heggen et~al.}(2005)\citenamefont{Heggen, Spaepen, and
  Feuerbacher}}]{HSF05}
\bibinfo{author}{\bibfnamefont{M.}~\bibnamefont{Heggen}},
  \bibinfo{author}{\bibfnamefont{F.}~\bibnamefont{Spaepen}}, \bibnamefont{and}
  \bibinfo{author}{\bibfnamefont{M.}~\bibnamefont{Feuerbacher}},
  \bibinfo{journal}{Journal Of Applied Physics} \textbf{\bibinfo{volume}{97}},
  \bibinfo{pages}{033506} (\bibinfo{year}{2005}).

\bibitem[{\citenamefont{Langer}(2006)}]{L06}
\bibinfo{author}{\bibfnamefont{J.~S.} \bibnamefont{Langer}},
  \bibinfo{journal}{Physical Review E} \textbf{\bibinfo{volume}{73}},
  \bibinfo{pages}{041504} (\bibinfo{year}{2006}).

\bibitem[{\citenamefont{Onuki}(2003)}]{O03}
\bibinfo{author}{\bibfnamefont{A.}~\bibnamefont{Onuki}},
  \bibinfo{journal}{Physical Review E} \textbf{\bibinfo{volume}{68}},
  \bibinfo{pages}{061502} (\bibinfo{year}{2003}).

\bibitem[{\citenamefont{Picard et~al.}(2002)\citenamefont{Picard, Ajdari,
  Bocquet, and Lequeux}}]{PABL02}
\bibinfo{author}{\bibfnamefont{G.}~\bibnamefont{Picard}},
  \bibinfo{author}{\bibfnamefont{A.}~\bibnamefont{Ajdari}},
  \bibinfo{author}{\bibfnamefont{L.}~\bibnamefont{Bocquet}}, \bibnamefont{and}
  \bibinfo{author}{\bibfnamefont{F.}~\bibnamefont{Lequeux}},
  \bibinfo{journal}{Physical Review E} \textbf{\bibinfo{volume}{66}},
  \bibinfo{pages}{051501} (\bibinfo{year}{2002}).

\bibitem[{\citenamefont{Argon and Kuo}(1979)}]{AK79}
\bibinfo{author}{\bibfnamefont{A.~S.} \bibnamefont{Argon}} \bibnamefont{and}
  \bibinfo{author}{\bibfnamefont{H.~Y.} \bibnamefont{Kuo}},
  \bibinfo{journal}{Materials Science And Engineering}
  \textbf{\bibinfo{volume}{39}}, \bibinfo{pages}{101} (\bibinfo{year}{1979}).

\bibitem[{\citenamefont{el~Kader and Earnshaw}(1999)}]{KE99}
\bibinfo{author}{\bibfnamefont{A.}~\bibnamefont{Abdel~Kader}} \bibnamefont{and}
  \bibinfo{author}{\bibfnamefont{J.~C.} \bibnamefont{Earnshaw}},
  \bibinfo{journal}{Physical Review Letters} \textbf{\bibinfo{volume}{82}},
  \bibinfo{pages}{2610} (\bibinfo{year}{1999}).

\bibitem[{\citenamefont{Dennin}(2004)}]{D04}
\bibinfo{author}{\bibfnamefont{M.}~\bibnamefont{Dennin}},
  \bibinfo{journal}{Physical Review E} \textbf{\bibinfo{volume}{70}},
  \bibinfo{pages}{041406} (\bibinfo{year}{2004}).

\bibitem[{\citenamefont{Durian}(1995)}]{D95}
\bibinfo{author}{\bibfnamefont{D.~J.} \bibnamefont{Durian}},
  \bibinfo{journal}{Phys. Rev. Lett.} \textbf{\bibinfo{volume}{75}},
  \bibinfo{pages}{4780} (\bibinfo{year}{1995}).

\bibitem[{\citenamefont{Lauridsen et~al.}(2002)\citenamefont{Lauridsen,
  Twardos, and Dennin}}]{LTD02}
\bibinfo{author}{\bibfnamefont{J.}~\bibnamefont{Lauridsen}},
  \bibinfo{author}{\bibfnamefont{M.}~\bibnamefont{Twardos}}, \bibnamefont{and}
  \bibinfo{author}{\bibfnamefont{M.}~\bibnamefont{Dennin}},
  \bibinfo{journal}{Physical Review Letters} \textbf{\bibinfo{volume}{89}},
  \bibinfo{pages}{098303} (\bibinfo{year}{2002}).

\bibitem[{\citenamefont{Gopal and Durian}(1995)}]{GD95}
\bibinfo{author}{\bibfnamefont{A.~D.} \bibnamefont{Gopal}} \bibnamefont{and}
  \bibinfo{author}{\bibfnamefont{D.~J.} \bibnamefont{Durian}},
  \bibinfo{journal}{Physical Review Letters} \textbf{\bibinfo{volume}{75}},
  \bibinfo{pages}{2610} (\bibinfo{year}{1995}).

\bibitem[{\citenamefont{Wang et~al.}(2007)\citenamefont{Wang, Krishan, and
  Dennin}}]{WKD07}
\bibinfo{author}{\bibfnamefont{Y.}~\bibnamefont{Wang}},
  \bibinfo{author}{\bibfnamefont{K.}~\bibnamefont{Krishan}}, \bibnamefont{and}
  \bibinfo{author}{\bibfnamefont{M.}~\bibnamefont{Dennin}},
  \bibinfo{journal}{Philosophical Magazine Letters}
  \textbf{\bibinfo{volume}{87}}, \bibinfo{pages}{125} (\bibinfo{year}{2007}).

\bibitem[{\citenamefont{Dennin and Knobler}(1997)}]{DK97}
\bibinfo{author}{\bibfnamefont{M.}~\bibnamefont{Dennin}} \bibnamefont{and}
  \bibinfo{author}{\bibfnamefont{C.~M.} \bibnamefont{Knobler}},
  \bibinfo{journal}{Physical Review Letters} \textbf{\bibinfo{volume}{78}},
  \bibinfo{pages}{2485} (\bibinfo{year}{1997}).

\bibitem[{\citenamefont{Dollet et~al.}(2006)\citenamefont{Dollet, Durth, and
  Graner}}]{DDG06}
\bibinfo{author}{\bibfnamefont{B.}~\bibnamefont{Dollet}},
  \bibinfo{author}{\bibfnamefont{M.}~\bibnamefont{Durth}}, \bibnamefont{and}
  \bibinfo{author}{\bibfnamefont{F.}~\bibnamefont{Graner}},
  \bibinfo{journal}{Physical Review E} \textbf{\bibinfo{volume}{73}},
  \bibinfo{pages}{061404} (\bibinfo{year}{2006}).

\bibitem[{\citenamefont{Reinelt and Kraynik}(2000)}]{RK00}
\bibinfo{author}{\bibfnamefont{D.~A.} \bibnamefont{Reinelt}} \bibnamefont{and}
  \bibinfo{author}{\bibfnamefont{A.~M.} \bibnamefont{Kraynik}},
  \bibinfo{journal}{J. Rheol. (N. Y.)} \textbf{\bibinfo{volume}{44}},
  \bibinfo{pages}{453} (\bibinfo{year}{2000}).

\bibitem[{\citenamefont{Vaz and Cox}(2005)}]{VC05}
\bibinfo{author}{\bibfnamefont{M.~F.} \bibnamefont{Vaz}} \bibnamefont{and}
  \bibinfo{author}{\bibfnamefont{S.~J.} \bibnamefont{Cox}},
  \bibinfo{journal}{Philosophical Magazine Letters}
  \textbf{\bibinfo{volume}{85}}, \bibinfo{pages}{415} (\bibinfo{year}{2005}).

\bibitem[{\citenamefont{Weaire et~al.}(1992)\citenamefont{Weaire, Bolton,
  Herdtle, and Aref}}]{WBHA92}
\bibinfo{author}{\bibfnamefont{D.}~\bibnamefont{Weaire}},
  \bibinfo{author}{\bibfnamefont{F.}~\bibnamefont{Bolton}},
  \bibinfo{author}{\bibfnamefont{T.}~\bibnamefont{Herdtle}}, \bibnamefont{and}
  \bibinfo{author}{\bibfnamefont{H.}~\bibnamefont{Aref}},
  \bibinfo{journal}{Phil. Mag. Lett.} \textbf{\bibinfo{volume}{66}},
  \bibinfo{pages}{293} (\bibinfo{year}{1992}).

\bibitem[{\citenamefont{Vincent-Bonnieu
  et~al.}(2006)\citenamefont{Vincent-Bonnieu, Hohler, and Cohen-Addad}}]{VHC06}
\bibinfo{author}{\bibfnamefont{S.}~\bibnamefont{Vincent-Bonnieu}},
  \bibinfo{author}{\bibfnamefont{R.~H.} \bibnamefont{Hohler}},
  \bibnamefont{and}
  \bibinfo{author}{\bibfnamefont{S.}~\bibnamefont{Cohen-Addad}},
  \bibinfo{journal}{Europhysics Letters} \textbf{\bibinfo{volume}{74}},
  \bibinfo{pages}{533} (\bibinfo{year}{2006}).

\bibitem[{\citenamefont{Pratt and Dennin}(2003)}]{pratt}
\bibinfo{author}{\bibfnamefont{E.}~\bibnamefont{Pratt}} \bibnamefont{and}
  \bibinfo{author}{\bibfnamefont{M.}~\bibnamefont{Dennin}},
  \bibinfo{journal}{Physical Review E} \textbf{\bibinfo{volume}{67}},
  \bibinfo{pages}{051402} (\bibinfo{year}{2003}).

\bibitem[{\citenamefont{Hutzler}(1997)}]{H97}
\bibinfo{author}{\bibfnamefont{S.}~\bibnamefont{Hutzler}}, Ph.D. thesis,
  \bibinfo{school}{Trinity College, Dublin} (\bibinfo{year}{1997}).

\bibitem[{\citenamefont{Lundberg et~al.}(2007)\citenamefont{Lundberg, Krishan,
  , Xu, O'Hern, and Dennin}}]{lkxod07b}
\bibinfo{author}{\bibfnamefont{M.}~\bibnamefont{Lundberg}},
  \bibinfo{author}{\bibfnamefont{K.}~\bibnamefont{Krishan}}, ,
  \bibinfo{author}{\bibfnamefont{N.}~\bibnamefont{Xu}},
  \bibinfo{author}{\bibfnamefont{C.~S.} \bibnamefont{O'Hern}},
  \bibnamefont{and} \bibinfo{author}{\bibfnamefont{M.}~\bibnamefont{Dennin}},
  \bibinfo{journal}{in preparation}  (\bibinfo{year}{2007}).

\bibitem[{\citenamefont{Allen and Tildesley}(1987)}]{allen}
\bibinfo{author}{\bibfnamefont{M.~P.} \bibnamefont{Allen}} \bibnamefont{and}
  \bibinfo{author}{\bibfnamefont{D.~J.} \bibnamefont{Tildesley}},
  \emph{\bibinfo{title}{Computer Simulation of Liquids}}
  (\bibinfo{publisher}{Oxford University Press}, \bibinfo{address}{Oxford},
  \bibinfo{year}{1987}).

\bibitem[{EPA()}]{EPAPS}
\bibinfo{note}{See EPAPS Document No. [] for movies of the reversible
  (ExperimentReversible.mov, SimulationReversible.mov) and irreversible
  (ExperimentIrreversible.mov, SimulationIrreversible.mov)events from both the
  experiments and simulation. For more information on EPAPS, see
  http://www.aip.org/pubservs/epaps.html.}

\bibitem[{\citenamefont{Khonik and Kosilov}(1994)}]{KK94}
\bibinfo{author}{\bibfnamefont{V.~A.} \bibnamefont{Khonik}} \bibnamefont{and}
  \bibinfo{author}{\bibfnamefont{A.~T.} \bibnamefont{Kosilov}},
  \bibinfo{journal}{Journal Of Non-Crystalline Solids}
  \textbf{\bibinfo{volume}{170}}, \bibinfo{pages}{270} (\bibinfo{year}{1994}).

\bibitem[{\citenamefont{Spaepen}(1977)}]{S77}
\bibinfo{author}{\bibfnamefont{F.}~\bibnamefont{Spaepen}},
  \bibinfo{journal}{Acta Metallurgica} \textbf{\bibinfo{volume}{25}},
  \bibinfo{pages}{407} (\bibinfo{year}{1977}).

\bibitem[{\citenamefont{Berthier}(2003)}]{B03}
\bibinfo{author}{\bibfnamefont{L.}~\bibnamefont{Berthier}},
  \bibinfo{journal}{Journal Of Physics-Condensed Matter}
  \textbf{\bibinfo{volume}{15}}, \bibinfo{pages}{S933} (\bibinfo{year}{2003}).

\bibitem[{\citenamefont{Picard et~al.}(2004)\citenamefont{Picard, Ajdari,
  Lequeux, and Bocquet}}]{PALB04}
\bibinfo{author}{\bibfnamefont{G.}~\bibnamefont{Picard}},
  \bibinfo{author}{\bibfnamefont{A.}~\bibnamefont{Ajdari}},
  \bibinfo{author}{\bibfnamefont{F.}~\bibnamefont{Lequeux}}, \bibnamefont{and}
  \bibinfo{author}{\bibfnamefont{L.}~\bibnamefont{Bocquet}},
  \bibinfo{journal}{European Physical Journal E} \textbf{\bibinfo{volume}{15}},
  \bibinfo{pages}{371} (\bibinfo{year}{2004}).

\end{thebibliography}

\end{document}